\documentclass[prd,reprint,showpacs,showkeys]{revtex4}
\usepackage{amsfonts}
\usepackage{amsthm}
\usepackage{amssymb}
\usepackage{amsmath}
\usepackage{graphicx}
\usepackage[font={footnotesize,it}]{caption}

\begin{document}

\title{Quantum Probe of Ho\v{r}ava-Lifshitz Gravity}
\author{O. Gurtug}
\affiliation{Maltepe University, Faculty of Engineering and Natural Sciences, Istanbul
-Turkey}
\affiliation{Department of Physics, Eastern Mediterranean University, G. Magusa, north
Cyprus, Mersin 10, Turkey. }
\email{ozaygurtug@maltepe.edu.tr;ozay.gurtug@emu.edu.tr}
\author{M. Mangut}
\email{161409101@st.maltepe.edu.tr}
\affiliation{Maltepe University, Faculty of Engineering and Natural Sciences, Istanbul
-Turkey}

\begin{abstract}
Particle probe analysis of the Kehagias - Sfetsos black hole spacetime of Ho%
\v{r}ava-Lifshitz gravity is extended to wave probe analysis within the
framework of quantum mechanics. The timelike naked singularity that develops
when $\omega M^{2}<1/2$, is probed with quantum fields obeying Klein-Gordon
and Chandrasekhar-Dirac equations. Quantum field probe of the naked
singularity has revealed that both the spatial part of the wave and the
Hamiltonian operators of Klein-Gordon and Chandrasekhar-Dirac equations are
essentially self-adjoint and thus, the naked singularity in the Kehagias -
Sfetsos spacetime become quantum mechanically non - singular.
\end{abstract}

\maketitle

\section{Introduction}

One of the most challenging problems of theoretical physics is how to merge
the physics occurring at small scales (quantum level) with those at large
scales (classical general relativity). The resolution to this problem is
extremely important, because at small scales, the classical general
relativity breaks down and the description of gravitational interaction
becomes impossible. The efforts toward constructing a\ consistent quantum
theory of gravity has encountered serious obstacles. One of these obstacles
is that Einstein's theory of classical general relativity (perturbatively)
is not a renormalizable theory and thus, the conventional quantization
techniques are not applicable.

However, alternative modified theories are developed to serve the resolution
of this problem. String \cite{1,2} and loop quantum gravity \cite{3}
theories are developed for dealing with the problems at small scales. It has
been shown in string theory that some timelike singularities are resolved.

In recent years, there has been a growing interest in another alternative
theory within the context of quantum gravity, $-$the Ho\v{r}ava - Lifshitz
(HL) theory of gravity \cite{4,5}. The HL theory incorporates an anisotropic
scaling of space and time. As a consequence of this scaling, while the
Lorentz invariance is broken at high energies (short distances, UV regime),
the Lorentz invariance is recovered at low energies (IR regime). The HL
gravity theory \textquotedblleft could therefore serve as a UV completion of
Einstein's general relativity\textquotedblright\ \cite{5}. The said theory
is also called in literature as the power-counting renormalizable theory.
Currently, there are several versions of HL theory that can be classified
whether or not the detailed balance and projectability conditions are
imposed.

In literature, there are a variety of studies related to the HL gravity. The
developments in this theory are collected and presented in a recent progress
report, prepared by Wang \cite{6}. Among the others, the spherically
symmetric solutions having characteristics analog to the Schwarzschild
solution have attracted considerable interest, in particular, the solution
found by Kahegias and Sfetsos (KS) \cite{7,8}, which describes static,
spherically symmetric black hole solution in the limiting case of $\left(
3+1\right) -$dimensional HL gravity. Whether in classical general relativity
or modified theories, solutions, admitting black holes are always more
fascinating and, as a result, attracts more attention. For example, the
underlying physics of the KS black hole solutions are extensively studied in
terms of geodesics (particle motion) \cite{9,10,11,12,13,14,15}. In addition
to aforementioned references, the bending of light and quasinormal modes of
KS black hole spacetime is also considered in \cite{16}.

However, solutions admitting naked singularities are always undervalued both
in classical general relativity and in modified theories. One reason that
may be linked to this view is the violation of Penrose's weak cosmic
censorship hypothesis (CCH). According to this hypothesis, all singularities
in physically realistic spacetimes are hidden by the horizons of black
holes, which preserve the deterministic nature of the classical general
relativity. However, many solutions have been found to Einstein's equations
that may exhibit naked singularities. The formation of the naked singularity
in the KS solution can be given as an example. And, the purpose of this
paper is to investigate this naked singularity within the framework of
quantum mechanics.

Spacetime singularities are predicted by Einstein's Theory of General
Relativity and are described as the \textit{geodesics incompleteness }with
respect to the point particle probe. Spacetime is geodesically incomplete,
if it contains at least one geodesic that is inextendible. In other words,
at the singularities, all the laws of physics are broken down, and that is
why, all these alternative theories are emerging to resolve this challenging
problem. The reasons for this are that before reaching the singularity, we
are in a microscopic region of the space that instead of the laws of
classical general relativity, the laws of quantum gravity are expected to be
replaced, hence, any attempt at investigating singularities in conjunction
with quantum mechanics should be considered as an important step in the
right direction.

In this study, the KS naked singular spacetime representing spherically
symmetric static vacuum solution in the HL gravity will be investigated
within the framework of quantum mechanics. The KS metric incorporates two
parameters, the gravitational mass parameter $M$ and the Ho\v{r}ava
parameter $\omega ,$ which represents the influence of the quantum effects.
These two parameters determine the physical characters of the KS spacetime.
If the product $\omega M^{2}\geq \frac{1}{2},$ the KS metric possesses a
black hole solution with two horizons. Thus, the curvature singularity at $%
r=0$ is covered by these horizons and preserves the CCH. However, if the
product $\omega M^{2}<\frac{1}{2}$, there are no horizons and the curvature
singularity at $r=0$ becomes visible to asymptotic observers, which is
called a naked singularity. \ The observational constraints on the value of $%
\omega $\ presented in \cite{12,17,18}, do not exclude the existence of the
KS naked singularities, hence, in light of these observational facts, it is
very important to focus on the naked singularities in the KS spacetimes. It
is shown in \cite{19} that the optical signatures of the KS naked
singularity is different from the signatures of the standard black holes in
classical general relativity. Furthermore, circular geodesics in the KS
naked singularity spacetimes are studied in \cite{20} and compared with the
counterparts in classical general relativity.

In this paper, we focus on the quantum nature of the KS naked singularity.
We investigate whether this classically singular spacetime remains quantum
mechanically singular or not. In our analysis, quantum particles (fields)
obeying the Klein-Gordon and the Dirac equations will be sent to the KS
naked singularity, thus, our analysis will be based on a wave probe, which
leads to the notion of \textit{quantum singularity}. In doing this, the work
of Wald \cite{21}, which was developed by Horowitz and Marolf (HM) \cite{22}
for static spacetimes will be used. The criterion of HM incorporates with
quantum field theory in curved spacetime. Hence, the analysis is based on
the motion of quantum particles (fields) in a classical curved background.

The paper is organized as follows. In section II, we give a brief review of
the KS spacetime and the description of the KS metric in a Newman-Penrose
formalism. In section III, the HM criterion is briefly explained. The naked
singularity in the KS spacetime is analyzed by probing the singularity with
quantum fields obeying the Klein-Gordon and Dirac equations. The paper ends
with a conclusion and discussion in section IV.

\section{Review of the KS Spacetime}

The $3+1-$dimensional action describing the Ho\v{r}ava - Lifshitz gravity is
given in \cite{5} as,%
\begin{eqnarray}
S &=&\int dtd^{3}x\sqrt{g}N\left\{ \frac{2}{\kappa ^{2}}\left(
K_{ij}K^{ij}-\lambda K^{2}\right) -\frac{\kappa ^{2}}{2\omega ^{4}}%
C_{ij}C^{ij}+\frac{\kappa ^{2}\mu }{2}\left( \frac{\epsilon ^{ijk}}{\omega
^{2}}R_{il}^{\left( 3\right) }\nabla _{j}R_{k}^{\left( 3\right) l}-\frac{%
R_{ij}^{\left( 3\right) }R^{\left( 3\right) ij}}{4}\right) \right. \\
&&\left. +\frac{\kappa ^{2}\mu }{8\left( 1-3\lambda \right) }\left( \frac{%
1-4\lambda }{4}\left( R^{\left( 3\right) }\right) ^{2}+\Lambda _{W}R^{\left(
3\right) }-3\Lambda _{W}^{2}\right) +\mu ^{4}R^{\left( 3\right) }\right\} , 
\notag
\end{eqnarray}%
in which 
\begin{equation}
K_{ij}=\frac{1}{2N}\left( \dot{g}_{ij}-\nabla _{i}N_{j}-\nabla
_{j}N_{i}\right) ,
\end{equation}%
is the second fundamental form and 
\begin{equation}
C^{ij}=\epsilon ^{ikl}\nabla _{k}\left( R_{l}^{\left( 3\right) j}-\frac{1}{4}%
R^{\left( 3\right) }\delta _{l}^{j}\right) ,
\end{equation}%
is the Cotton tensor with, $\kappa ,\lambda $ and $\omega $ as dimensionless
coupling constants. On the other hand, $\mu $ and $\Lambda _{W}$ are
dimensionful with mass $[\mu ]=1$ and $[\Lambda _{W}]=2.$ The corresponding
metric is given by%
\begin{equation}
ds^{2}=-N^{2}dt^{2}+g_{ij}\left( dx^{i}+N^{i}dt\right) \left(
dx^{j}+N^{j}dt\right) ,
\end{equation}%
where $g_{ij},N^{i}$ and $N$ are the dynamical fields.

The above Lagrangian has been considered by Kehagias and Sfetsos in the
limiting case of $\Lambda _{W}\rightarrow 0$ and $\lambda =1,$ which leads
to, perhaps, one of the important solution obtained so far within the
context of Ho\v{r}ava - Lifshitz gravity. The obtained solution is
spherically symmetric, which describes asymptotically flat black hole
metric. This metric is important, because, it constitutes the analog of
Schwarzschild solution of classical general relativity. \ The metric
obtained by KS is given by%
\begin{equation}
ds^{2}=-f(r)dt^{2}+\frac{dr^{2}}{f(r)}+r^{2}\left( d\theta ^{2}+\sin
^{2}\theta d\varphi ^{2}\right) ,
\end{equation}%
in which 
\begin{equation}
N^{2}=f(r)=1+\omega r^{2}-\sqrt{r\left( \omega ^{2}r^{3}+4\omega M\right) },
\end{equation}%
where $M$ is an integration constant with dimension $[M]=-1$ and $\omega
=16\mu ^{2}/\kappa ^{2}.$ The obtained metric possesses black hole solution
with two horizons located at 
\begin{equation}
r_{\pm }=M\left( 1\pm \sqrt{1-\frac{1}{2\omega M^{2}}}\right) ,
\end{equation}%
provided that $\omega M^{2}\geq \frac{1}{2}.$ The Ricci scalar diverges as $%
1/r^{3/2},$ indicating true curvature singularity at $r=0$, which is covered
by horizons. The metric (5), has interesting properties that for large $r$
in fixed $\omega $ or large $\omega $ in fixed $r$, possesses usual
Schwarzschild black hole behavior. This is the case whenever $r\gg \left(
M/\omega \right) ^{1/3}$, which allows the metric function (6) to be written
in the following form, \ 
\begin{equation}
f(r)\approx 1-\frac{2M}{r}+\mathcal{O}\left( r^{-4}\right) .
\end{equation}%
The classical properties of this particular case has been analyzed in \cite%
{15}, by investigating particle geodesics. It is shown in the analysis that
the KS black hole solution is more flattened compared with Schwarzschild
black hole in and around horizon, and as a result of this effect, the
gravitation becomes weaker near the center of the region. \ The effective
potential of the KS black hole, as in the case of Reissner-Nordstr\"{o}m
black hole, has a repulsive character. The overall effect of these
properties on the particle motion is that no particle falls to the center
like Schwarzschild black hole, but particles are scattered to infinity or
trapped in periodic orbits. This behavior in the particle motion is
persistent with non - vanishing angular momentum \cite{13}. But, for the
radial motion (with vanishing angular momentum), null geodesics always reach
the curvature singularity at $r=0$, while timelike geodesics depending on
their energy $E$ are either trapped on a radial geodesics or reach the
singularity at $r=0$. This picture can be made more transparent in the
following way. For $r>r_{+},$ the coordinate $r$ is a spacelike, $t$ is a
timelike and the coordinate singularity at $r=r_{+}$ is interpreted as event
(outer) horizon. Inside this event horizon $r_{-}<r<r_{+},$ $r$ is timelike, 
$t$ is spacelike and the coordinate singularity at $r=r_{-},$ is interpreted
as the inner horizon. However, in the innermost region $0<r<r_{-},$ the
space is static and the coordinate $r$ becomes spacelike, $t$ becomes
timelike. In addition, the character of the curvature singularity at $r=0$
is timelike. The timelike character of the singularity in spherically
symmetric case can be seen by analysing the behavior of tortoise coordinate $%
r_{\ast }$ in the limit of $r\rightarrow 0.$ If the singularity is at a
finite value of $r_{\ast }$, then it is timelike, but if it is at $r_{\ast
}=-\infty ,$ the singularity is null. In the case of KS black hole solution,
the tortoise coordinate is given by, 
\begin{equation}
r_{\ast }=r+\frac{1}{2\kappa _{+}}\ln \left\vert \frac{r-r_{+}}{r_{+}}%
\right\vert +\frac{1}{2\kappa _{-}}\ln \left\vert \frac{r-r_{-}}{r_{-}}%
\right\vert ,
\end{equation}%
in which $\kappa _{\pm }=\frac{2\omega r_{\pm }^{2}-1}{4r_{\pm }\left(
\omega r_{\pm }^{2}+1-\right) }.$ The limit $r\rightarrow 0,$ reveals that $%
r_{\ast }=0$ and hence, it is timelike. This structure of KS black hole is
in marked contrast when compared with Schwarzchild black hole whose
innermost region is dynamic and the curvature singularity at $r=0$ is
spacelike.

The KS naked singularity possesses similar behaviour against a particle
probe. We show this by calculating the geodesic equations in the naked
singular KS spacetime. The KS solution becomes naked singular whenever $%
\omega M^{2}<\frac{1}{2}.$ In this paper, we will take $M=\frac{1}{2}$ and $%
\omega =1$, which satisfies the naked singularity condition and the
corresponding metric function near the singularity is given in Eq.(30). The
conserved quantities are the energy $E$ and the angular momentum $l.$
Restricting the motion in an equatorial plane $\theta =\pi /2$, we have the
following constants of motion,%
\begin{equation}
\frac{dt}{d\tau }=-\frac{E}{f(r)},\text{ \ \ \ \ \ \ \ \ \ \ \ \ \ \ }\frac{%
d\varphi }{d\tau }=\frac{l}{r^{2}}.
\end{equation}%
Using these conserved quantities we obtain%
\begin{equation}
\left( \frac{dr}{d\tau }\right) ^{2}=E^{2}-f(r)\left( \epsilon +\frac{l^{2}}{%
r^{2}}\right) ,
\end{equation}%
and%
\begin{equation}
\left( \frac{dr}{d\varphi }\right) ^{2}=\frac{2r^{4}}{l^{2}}\left(
\varepsilon _{eff}-V_{eff}\left( r\right) \right) ,
\end{equation}%
in which $\tau $ is a proper time, the effective potential $V_{eff}\left(
r\right) $ \ and the effective energy $\varepsilon _{eff}$ are given by, 
\begin{equation}
V_{eff}\left( r\right) =\frac{1}{2}f(r)\left( \epsilon +\frac{l^{2}}{r^{2}}%
\right) ,\text{ \ \ \ \ \ \ }\varepsilon _{eff}=\frac{E^{2}}{2}.\text{\ \ }
\end{equation}

We consider motion with zero angular momentum (i.e. $l=0$) such that the
particles move radially. Radial null geodesics ($\epsilon =0$), imply the
motion of massless particle (photon). The time required for a photon to
reach a singularity from an initial position $r_{0}$ can be calculated using
Eq.(11) and is given by

\begin{equation}
\pm (t-t_{0})=\sqrt{2r_{0}}+\ln \left\vert 1-\sqrt{2r_{0}}\right\vert ,
\end{equation}%
here, $t$ is the time measured by a distant observer and $t_{0}$ is the
initial time. Similar calculation for the timelike geodesics ($\epsilon =1$%
), which defines the motion of massive particles yields,%
\begin{equation}
\pm (t-t_{0})=-\frac{4}{3}\delta ^{3/2}+2\delta \sqrt{\delta +\sqrt{2r_{0}}}-%
\frac{2}{3}\left( \delta +\sqrt{2r_{0}}\right) ^{3/2},
\end{equation}%
in which $\delta =E^{2}-1.$ Using Eq.(11), we also calculate the radial
acceleration acting on the massive particle for $l=0$ case. This calculation
reveals that%
\begin{equation}
\frac{d^{2}r}{d\tau ^{2}}=-\frac{1}{2}\frac{df(r)}{dr}=\frac{1}{\sqrt{8r}}>0.
\end{equation}%
Positive radial acceleration implies repulsive force on the particle that
follows timelike geodesics. As a result, depending on the energy $E$ of the
particle, this repulsive force may reflect the particle back.

\subsection{The Description of the KS solution in a Newman-Penrose Formalism}

The KS metric is investigated with the Newman-Penrose (NP) formalism, in
order to clarify the contribution of the parameter $\omega $. \ The set of
proper null tetrads $1-forms$ is given by%
\begin{eqnarray}
l &=&dt-\frac{dr}{f(r)}, \\
n &=&\frac{1}{2}\left( f(r)dt+dr\right) ,  \notag \\
m &=&-\frac{r}{\sqrt{2}}\left( d\theta +i\sin \theta d\varphi \right) , 
\notag \\
\bar{m} &=&-\frac{r}{\sqrt{2}}\left( d\theta -i\sin \theta d\varphi \right) .
\notag
\end{eqnarray}%
The non-zero spin coefficients in this tetrad are%
\begin{equation}
\beta =-\alpha =\frac{\cot \theta }{2\sqrt{2}r},\text{ \ \ }\rho =-\frac{1}{r%
},\text{ \ \ }\mu =-\frac{f(r)}{2r},\text{ \ \ }\gamma =\frac{1}{4}\frac{%
df(r)}{dr}.
\end{equation}%
The non - zero Weyl and the Ricci scalars are%
\begin{equation}
\Psi _{2}=-\frac{M}{r^{3}}\left( 1+\frac{4M}{\omega r^{3}}\right) ^{-1/2},
\end{equation}%
\begin{equation}
\phi _{11}=\frac{9M^{2}}{2r^{6}\omega }\left( 1+\frac{4M}{\omega r^{3}}%
\right) ^{-3/2},
\end{equation}%
\begin{equation}
\Lambda =-\frac{\omega }{2}+\frac{\omega }{2}\left( 1+\frac{4M}{\omega r^{3}}%
\right) ^{-1/2}+\frac{M}{r^{3}}\left( 1+\frac{4M}{\omega r^{3}}\right)
^{-3/2}+\frac{5M^{2}}{2r^{6}\omega }\left( 1+\frac{4M}{\omega r^{3}}\right)
^{-3/2}.
\end{equation}%
The spacetime character is Petrov $type-D,$ since, the only non - zero Weyl
scalar is $\Psi _{2}$. The parameter $\omega $ represents the contribution
of HL gravity. \ In the large limit of $\omega \gg 1,$ the Ricci scalars $%
\phi _{11}$ and $\Lambda $ vanishes, and the Weyl scalar $\Psi _{2}$ remains
the only non - zero component with a value of $\Psi _{2}\approx -\frac{M}{%
r^{3}},$ as in the case of Schwarzschild black hole.

\section{Quantum Probe of the KS Naked Singularity}

As explained with justifications in the introduction, the main purpose of
this paper is to analyze the naked singularity of the KS spacetime with
quantum particles/fields. To serve the purpose, the criterion developed by
HM will be used in this study. According to this criterion; the classically
singular spacetime remains quantum mechanically singular, if the spatial
part of the wave operator is not essentially self-adjoint. If this is the
case, then, the future time-evolution is not uniquely determined and hence,
the corresponding spacetime is regarded as quantum mechanically singular or
quantum singular. Thus, the HM criterion requires a unique time-evolution in
order to say that the corresponding spacetime is quantum mechanically
regular or quantum regular.

The general mathematical formalism of this criterion is given in detail in 
\cite{23,24,25} . At this stage, we prefer to give the main theme of the HM
criterion. \ Let us consider a relativistic scalar particle/field with mass $%
\tilde{m}$ satisfying the Klein-Gordon equation. The key point is to split
the spatial and time part of the Klein- Gordon equation and write it in the
form of%
\begin{equation}
\frac{\partial ^{2}\psi }{\partial t^{2}}=-A\psi ,
\end{equation}%
where $A$ is the spatial wave operator. According to the HM criterion, the
singular character of the spacetime with respect to quantum probe is
characterized by investigating whether the spatial part of the wave operator 
$A$ has unique self - adjoint extensions (i.e. essentially self - adjoint)
in the entire space or not. If the extension is unique, it is said that the
space is quantum mechanically regular. In order to make this point more
clear, consider the Klein- Gordon equation for a free particle that satisfies%
\begin{equation}
i\frac{d\psi }{dt}=\sqrt{A_{E}}\psi ,
\end{equation}%
whose solution is%
\begin{equation}
\psi \left( t\right) =e^{-it\sqrt{A_{E}}}\psi \left( 0\right) ,
\end{equation}%
in which $A_{E}$ denotes the extension of the wave operator $A$. If $A$ does
not have unique self - adjoint extensions, then the future time evolution of
the wave function (24) is ambiguous. And, HM criterion defines the spacetime
as quantum mechanically singular. But, if the wave operator $A$ \ has a
unique self-adjoint extension, then the future time evolution of the quantum
particle described by (24) is uniquely determined by the initial conditions
and the criterion of HM, defines this spacetime as quantum mechanically
regular.

More specifically, the HM criterion incorporates with the self-adjointness
of the operators. Thus, the space on which these operators operate must be
specified. The natural function space for quantum mechanics is the Hilbert
space, which is known to be the space of square integrable functions $%
L^{2}\left( 0,\infty \right) .$ The characteristic property of the Hilbert
space is that the squared norm of the solution (let's say $R$) associated
with the operator must satisfy the condition of $\mathcal{H}=\left\{
R:\left\Vert R\right\Vert ^{2}\text{ }<\infty \right\} .$ In fact, the
condition on the squared norm $\left\Vert R\right\Vert ^{2}$ $<\infty ,$
implies that the solution \textit{exists }and does not include \textit{%
unbounded} solutions. It is also important to note that, the operator $A$ is
a symmetric and positive operator on the Hilbert space $\mathcal{H}$ \cite%
{22}

To test for essential self-adjointness, two powerful methods, namely, the
standard method of von Neumann's deficiency indices and the Weyl's limit
point - limit circle criterion (used in \cite{31,32,33,34}, for the same
purpose ) are the most widely used ones. In this paper, both methods will be
used for investigating the self-adjointness of the wave operator $A$. The
mathematical definitions of these two methods are given in Appendix to
ensure the reader's comprehensive understanding.

In this study, the naked singularity of the KS spacetime will be probed with
two different quantum particles: $spin-0$ scalar particles obeying the
Klein-Gordon equation and $spin-1/2$ particles obeying the Dirac equation.

\subsection{The Klein-Gordon fields}

The massive Klein-Gordon equation in general is given by,%
\begin{equation}
\left( \frac{1}{\sqrt{-g}}\partial _{\mu }\left[ \sqrt{g}g^{\mu \nu
}\partial _{\nu }\right] -\tilde{m}^{2}\right) \psi =0,
\end{equation}%
in which $\tilde{m}$ is the mass of the scalar particle. The Klein-Gordon
equation is written for the metric (5) and after separating time and spatial
parts, we have

\begin{eqnarray}
\frac{\partial ^{2}\psi }{\partial t^{2}} &=&f^{2}\left( r\right) \frac{%
\partial ^{2}\psi }{\partial r^{2}}+\frac{f\left( r\right) }{r^{2}}\frac{%
\partial ^{2}\psi }{\partial \theta ^{2}}+\frac{f(r)}{r^{2}\sin ^{2}\theta }%
\frac{\partial ^{2}\psi }{\partial \varphi ^{2}}+\frac{f(r)\cot \theta }{%
r^{2}}\frac{\partial \psi }{\partial \theta }+f(r)\left( \frac{2f\left(
r\right) }{r}+f^{\prime }\left( r\right) \right) \frac{\partial \psi }{%
\partial r} \\
&&-f\left( r\right) \tilde{m}^{2}\psi .  \notag
\end{eqnarray}%
When we compare equations (26) and (22), the spatial part of the wave
operator is written as

\begin{equation*}
\emph{A}=-f^{2}\left( r\right) \frac{\partial ^{2}}{\partial r^{2}}-\frac{%
f\left( r\right) }{r^{2}}\frac{\partial ^{2}}{\partial \theta ^{2}}-\frac{%
f(r)}{r^{2}\sin ^{2}\theta }\frac{\partial ^{2}}{\partial \varphi ^{2}}-%
\frac{f(r)\cot \theta }{r^{2}}\frac{\partial }{\partial \theta }-f(r)\left( 
\frac{2f\left( r\right) }{r}+f^{\prime }\left( r\right) \right) \frac{%
\partial }{\partial r}+f\left( r\right) \tilde{m}^{2}
\end{equation*}

The next step is to test the spatial part of the wave operator \emph{A} for
essential self-adjointness.

\subsubsection{Method 1: The von Neumann Criterion of Deficiency Indices}

The standard method that deals with the concept of deficiency indices was
discovered by Weyl \cite{26} and generalized by von Neumann \cite{27} in the
Theorem 1 given in Appendix. The determination of the deficiency indices $%
\left( n_{+},n_{-}\right) $ of the operator $A$, is reduced to count the
number of solutions to equation%
\begin{equation}
\left( A^{\ast }\pm i\right) \psi =0,
\end{equation}%
that belongs to the Hilbert space $\mathcal{H}.$ If there are no square
integrable $(L^{2}\left( 0,\infty \right) )$ solutions (i.e., $n_{+}=n_{-}=0$%
) in the entire space, the operator $A$ possesses a unique self-adjoint
extension and it is called essentially self-adjoint. Consequently, the
method to find a sufficient condition for the operator $A$ to be essentially
self-adjoint is to investigate the solutions satisfying equation (27)\ that
do not belong to the Hilbert space $\mathcal{H}$.

Applying separation of variables to equation (27), in the form of $\psi
=R\left( r\right) Y_{l}^{m}\left( \theta ,\varphi \right) ,$ yields the
following radial equation for $R(r)$:

\begin{equation}
R^{\prime \prime }+\frac{\left( r^{2}f\right) ^{\prime }}{fr^{2}}R^{\prime }-%
\left[ \frac{l\left( l+1\right) }{fr^{2}}+\tilde{m}^{2}\pm \frac{i}{f^{2}}%
\right] R=0,
\end{equation}%
in which prime denotes the derivative with respect to $r$ and $R=R(r)$.

The square integrability of the solutions of (28) for each sign $\pm $ is
checked by calculating the squared norm, in which the function space on each 
$t=$ constant hypersurface $\Sigma _{t}$ is defined as $\mathcal{H}=\left\{
R:\parallel R\parallel ,exist\text{ and }finite\right\} .$ The squared norm
for $\left( 3+1\right) $ $-$dimensional space can be defined as \cite{22},%
\begin{equation}
\Vert R\Vert ^{2}=\int_{\Sigma _{t}}\sqrt{-g}g^{tt}RR^{\ast }d^{3}\Sigma
_{t}.
\end{equation}%
The spatial operator $A$ is essentially self-adjoint if neither of the
solutions of Eq.(28) is square integrable over all space $L^{2}\left(
0,\infty \right) .$ The behavior of the Eq.(28), near $r\rightarrow 0$ and $%
r\rightarrow \infty $ will be considered separately in the following
subsections.

Since our aim is to analyze the naked singularity of the KS spacetime, it is
important to note that in our analysis, the mass parameter $M$ and the Ho%
\v{r}ava parameter $\omega $ will be chosen in such a way that the
inequality $\omega M^{2}<\frac{1}{2}$ holds. Therefore, if $M=\frac{1}{2},$
then the Ho\v{r}ava parameter $\omega <2.$ In the rest of the paper, the
mass parameter and the Ho\v{r}ava parameter are taken as $M=\frac{1}{2}$ and 
$\omega =1,$ respectively.

\paragraph{The case of $r\rightarrow 0:$}

In the case when $r\rightarrow 0,$ the metric function (6) behave as 
\begin{equation}
f(r)\approx 1-\sqrt{2r}+\mathcal{O}\left( r^{2}\right) ,
\end{equation}%
thus, the Eq.(28) is simplified to,

\begin{equation}
R^{\prime \prime }+\frac{9}{2r}R^{\prime }-\frac{l\left( l+1\right) }{r^{2}}%
R=0,
\end{equation}%
whose solution is%
\begin{equation}
R(r)=C_{1}r^{\gamma _{1}}+C_{2}r^{\gamma _{2}},
\end{equation}%
in which $C_{1},$ $C_{2}$ are the integration constants and 
\begin{equation}
\gamma _{1}=\frac{1}{4}\left( -7+\sqrt{49+16l\left( l+1\right) }\right) ,%
\text{ \ \ \ \ \ \ \ \ \ \ }\gamma _{2}=-\frac{1}{4}\left( 7+\sqrt{%
49+16l\left( l+1\right) }\right) .\text{\ \ \ }
\end{equation}

The square integrability of the solution (32) is checked by calculating the
squared norm defined in equation (29) in the limiting case of the metric (5)
when $r\rightarrow 0,$ which is given by%
\begin{equation}
\Vert R\Vert ^{2}\sim \int_{0}^{const.}\frac{r^{2}\left\vert R\right\vert
^{2}}{\left( 1-\sqrt{2r}\right) }dr.
\end{equation}%
We perform the analysis for different modes of solution. If $l=0$, which
corresponds to \textit{s-wave} mode, the solution becomes $R(r)=C_{1}+\frac{%
C_{2}}{r^{7/2}}.$ The square integrability analysis for this particular
solution has revealed that $\Vert R\Vert ^{2}\rightarrow \infty $, which is
not square integrable, thus, the solution does not belong to the Hilbert
space. If $l\neq 0,$ as long as $C_{1}=0$ and $C_{2}\neq 0,$ the square
integrability condition indicates that $\Vert R\Vert ^{2}\rightarrow \infty
, $ hence the solution does not belong to Hilbert space.

\paragraph{The case of $r\rightarrow \infty :$}

In the case when $r\rightarrow \infty ,$ the metric function (6) behave as 
\begin{equation}
f(r)\approx 1-\frac{1}{r}+\mathcal{O}\left( r^{-4}\right) .
\end{equation}%
Thus, the Eq.(28) reduces to%
\begin{equation}
R^{\prime \prime }+\frac{2}{r}R^{\prime }+\left( -\tilde{m}^{2}\pm i\right)
R=0,
\end{equation}%
whose solution is given by%
\begin{equation}
R(r)=\frac{C_{3}}{r}\sin \kappa r+\frac{C_{4}}{r}\cos \kappa r,
\end{equation}%
in which $\kappa =\sqrt{\pm i-\tilde{m}^{2}},$ and $C_{3},$ $C_{4}$ are the
integration constants ( in general complex). The square integrability is
checked with the following norm written for the case $r\rightarrow \infty ,$%
\begin{equation}
\Vert R\Vert ^{2}\sim \int_{const.}^{\infty }\frac{r^{3}\left\vert
R\right\vert ^{2}}{r-1}dr.
\end{equation}%
When the solution (37) is substituted into equation (38), with $%
C_{3}=C_{4}=1,$ we have the following integral to be integrated%
\begin{equation}
\Vert R\Vert ^{2}\sim \int_{const}^{\infty }\left( \frac{r}{r-1}\right)
\left( 1+2\sin \kappa r\cos \kappa r\right) dr=\int_{const}^{\infty }\frac{%
rdr}{r-1}+2\int_{const}^{\infty }\left( \frac{r\sin \kappa r\cos \kappa r}{%
r-1}\right) dr.
\end{equation}%
The first integral can be integrated easily and the result is 
\begin{equation}
\int_{const}^{\infty }\frac{rdr}{r-1}=\left( r-1+\ln \left\vert
r-1\right\vert \right) \left\vert _{const}^{\infty }\right. \rightarrow
\infty .
\end{equation}%
The second integral is evaluated by using the comparison test, especially
developed for the improper integrals. The second integral can be written as, 
\begin{equation}
I=\int_{const}^{\infty }\left( \frac{r2\sin \kappa r\cos \kappa r}{r-1}%
\right) dr=\int_{const}^{\infty }\left( \frac{r\sin \left( 2\kappa r\right) 
}{r-1}\right) dr
\end{equation}%
We replace $\sin \left( 2\kappa r\right) $ with its power series expansion,%
\begin{equation}
\sin \left( 2\kappa r\right) =\sum_{n=0}^{\infty }\left( -1\right) ^{n}\frac{%
\left( 2\kappa r\right) ^{2n+1}}{\left( 2n+1\right) !},
\end{equation}%
and the second integral becomes%
\begin{equation}
I=\int_{const}^{\infty }\left( \frac{r}{r-1}\right) \left\{
\sum_{n=0}^{\infty }\left( -1\right) ^{n}\frac{\left( 2\kappa r\right)
^{2n+1}}{\left( 2n+1\right) !}\right\} dr=\sum_{n=0}^{\infty }\left(
-1\right) ^{n}\frac{\left( 2\kappa \right) ^{2n+1}}{\left( 2n+\right) !}%
\int_{const}^{\infty }\left( \frac{r^{a}}{r-1}\right) dr,
\end{equation}%
in which $a=2n+2$. It should be noted that the series in front of the second
integral is analysed with D'Alambert ratio test for a convergency. It is
found that the series is absolute convergent. If we let $t=r-1;$ since $r\gg
1,$ implies $t\gg 1,$ and the second integral becomes proportional to the
following integral%
\begin{equation}
\sim \int_{c}^{\infty }\frac{\left( t+1\right) ^{a}}{t}dt.
\end{equation}%
As a requirement of the comparison test, we define the following inequality,%
\begin{equation}
0\leq \frac{t+1}{t}\leq \frac{\left( t+1\right) ^{a}}{t}.
\end{equation}%
The following integral can be evaluated easily and we find that it diverges, 
\begin{equation}
\int_{c}^{\infty }\left( \frac{t+1}{t}\right) dt=\left( t+\ln \left\vert
t\right\vert \right) \left\vert _{const}^{\infty }\right. \rightarrow \infty
.
\end{equation}%
According to the comparison test, divergence of the integral $%
\int_{c}^{\infty }\left( \frac{t+1}{t}\right) dt,$ implies the divergence of 
$\int_{c}^{\infty }\frac{\left( t+1\right) ^{a}}{t}dt$ . In view of this
analysis, the solution (37) fails to satisfy square integrability condition,
and hence, does not belong to the Hilbert space.

The method of von Neumann's deficiency theorem for defining whether the
operator $A$ has a unique self-adjoint extension (or essentially
self-adjoint) or not, necessitates the investigation of the solution of
Eq.(28) in the entire space $\left( 0,\infty \right) $ and the counting of
the number of solutions that does not belong to the Hilbert space. In other
words, if there is one solution that fails to be square integrable for the
entire space, then the operator $A$ is said to be essentially self-adjoint.
Our analysis has shown that the solutions of the Eq. (28), near $%
r\rightarrow 0$ and $r\rightarrow \infty ,$ are not square integrable.
Hence, the operator $A$ is essentially self-adjoint and the future time
evolution of the quantum particles/waves can be predicted uniquely.
Consequently, the classical naked singularity in the KS spacetime becomes
quantum mechanically regular when probed with massive bosons described by
the Klein-Gordon equation.

\subsubsection{Method 2: Weyl's limit circle - limit point criterion}

The massive Klein-Gordon equation given in Eq.(25) has mode solutions in the
following separable form 
\begin{equation}
\psi =\frac{1}{r}e^{-i\widetilde{\omega }t}R(r)Y_{lm}(\theta ,\varphi ),
\end{equation}%
where $\widetilde{\omega }$ is the frequency of the scalar wave, $%
Y_{lm}(\theta ,\varphi )$ are spherical harmonics and $r$ is the radial
coordinate. The radial part of the wave equation is obtained as%
\begin{equation}
R^{\prime \prime }+\frac{f^{\prime }}{f}R^{\prime }+\frac{1}{f}\left[ \frac{%
f^{\prime }}{f}-\frac{l\left( l+1\right) }{r^{2}}-\tilde{m}^{2}+\frac{%
\widetilde{\omega }^{2}}{f}\right] R=0,
\end{equation}%
where $l$ is a separation constant. In this method, one has to write the
equation (48) in one-dimensional Schr\"{o}dinger - like wave equation and
investigate its effective potential near $r\rightarrow \infty $ and $%
r\rightarrow 0$.

\paragraph{The case when $r\rightarrow \infty :$}

In order to write the above equation in one-dimensional Schr\"{o}dinger -
like wave equation, we use the tortoise coordinates defined by $dr_{\ast }=%
\frac{dr}{f}$ and we found that 
\begin{equation}
r_{\ast }=r+\ln \left\vert r-1\right\vert .
\end{equation}%
Note that in this particular limit the metric function given in equation
(35) is used. In tortoise coordinates the radial wave equation (48) becomes 
\begin{equation}
\frac{d^{2}R}{dr_{\ast }^{2}}-\left( \frac{r-1}{r}\right) \left[ \frac{1}{%
r^{3}}-\frac{l\left( l+1\right) }{r^{2}}-\tilde{m}^{2}+\frac{r\widetilde{%
\omega }^{2}}{r-1}\right] R=0.
\end{equation}%
It should be noted that the above equation involves two variables $r$ and $%
r_{\ast }$. It must be reduced to a single variable. To do this, we use the
standard logarithmic inequality defined by $\ln \left( x\right) \leq x-1;$ \
for $x>0$. In our case $x=r-1$, hence, the logarithmic inequality becomes, $%
\ln \left( r-1\right) \leq r-2;$ \ \ for \ $r>1$.\ This inequality leads us
to state, $r>r-2>\ln \left( r-1\right) $. Since we are interested as $%
r\rightarrow \infty ,$ then $r\gg \ln \left\vert r-1\right\vert $ and the
tortoise coordinate approximates to $r_{\ast }\simeq r.$ From this result, $%
\frac{d^{2}R}{dr_{\ast }^{2}}=\frac{d^{2}R}{dr^{2}}$ in equation (50). So,
the equation (50) can be written as function of $r$ only in the following
form%
\begin{equation}
\frac{d^{2}R}{dr^{2}}+\left[ -\widetilde{\omega }^{2}+V(r)\right] R=0,
\end{equation}%
in which $V(r)$ is the effective potential given by%
\begin{equation}
V(r)=\frac{l\left( l+1\right) \left( r-1\right) }{r^{3}}+\frac{\left(
r-1\right) \tilde{m}^{2}}{r}-\frac{r-1}{r^{4}}.
\end{equation}%
Now, we will apply the Weyl's limit circle-limit point criterion. Note that
the potential $V(r)$ is bounded below such that,%
\begin{equation}
V(r)\geq -\left\{ \frac{1+l\left( l+1\right) }{r^{3}}+\frac{\tilde{m}^{2}}{r}%
\right\} .
\end{equation}%
As a requirement of the Theorem 3 item (i), we define the positive
differentiable function $M(r)$ as,%
\begin{equation}
M(r)=\frac{1+l\left( l+1\right) }{r^{3}}+\frac{\tilde{m}^{2}}{r},
\end{equation}%
and item (ii) of Theorem 3 imposes the condition that the integration of $%
\left( M(r)\right) ^{-1/2}$ must be, 
\begin{equation}
\int_{1}^{\infty }\left( M(r)\right) ^{-1/2}dr=\infty .
\end{equation}%
Our calculation has revealed that the requirement (ii) of Theorem 3 is
satisfied. Finally, the last condition which states that $\left( M(r)\right)
^{\prime }/\left( M(r)\right) ^{3/2}$ is bounded near $\infty $, is verified
by calculating $\left( M(r)\right) ^{\prime }/\left( M(r)\right) ^{3/2}$
explicitly and given by%
\begin{equation}
\left( M(r)\right) ^{\prime }/\left( M(r)\right) ^{3/2}=\frac{1}{\sqrt{\frac{%
1+l(l+1)}{r}+\tilde{m}^{2}r}}.
\end{equation}%
It is clear that in the limit $r\rightarrow \infty $, this resulting
expression is bounded near $\infty .$ This analysis shows that the effective
potential near $\infty $ is in the limit point case. As a result, the
Hamiltonian operator has a unique extension and thus, it is essentially
self-adjoint.

\paragraph{The case when $r\rightarrow 0:$}

In the case when $r\rightarrow 0,$ the metric function (30) is used. The
tortoise coordinate \ and the corresponding one-dimensional Schr\"{o}dinger
- like wave equation reads as

\begin{equation}
r_{\ast }=1-\sqrt{2r}-\ln \left\vert 1-\sqrt{2r}\right\vert
\end{equation}

\begin{equation}
-\frac{d^{2}R}{dr_{\ast }^{2}}+V(r)R=-\widetilde{\omega }^{2}R
\end{equation}%
in which the effective potential $V(r)$ is given by%
\begin{equation}
V(r)=\frac{l\left( l+1\right) \left( 1-\sqrt{2r}\right) }{r^{2}}+\frac{1}{%
\sqrt{2r}}+\tilde{m}^{2}\left( 1-\sqrt{2r}\right) .
\end{equation}%
As before, one-dimensional Schr\"{o}dinger - like wave equation (58)
involves two variables $r$ and $r_{\ast }$. It must be reduced to a single
variable. Recall, the series expansion of logarithmic function, $\ln \left(
1\pm x\right) =\pm x-\frac{1}{2}x^{2}\pm \frac{1}{3}x^{3}...,$ \ and using
it in the equation (57), we obtain that the tortoise coordinate approximates
to $r_{\ast }\simeq 1+r.$ Hence, $\frac{d^{2}R}{dr_{\ast }^{2}}=\frac{d^{2}R%
}{dr^{2}}$ in equation (58) and becomes as a function of $r$ only. We are
interested in the leading behavior of the effective potential as $%
r\rightarrow 0.$ Thus, the first term is the dominant term of the effective
potential and in this particular limit one has,%
\begin{equation}
V(r)\sim \frac{l\left( l+1\right) }{r^{2}}.
\end{equation}%
Thus, by Theorem 4, if $l\left( l+1\right) \geq 3/4$ then the Hamiltonian
operator is in the limit point case at zero and therefore, it is essentially
self-adjoint.

We can thus conclude that the classical KS naked singular spacetime is
quantum mechanically non - singular. This is proved by analyzing the
self-adjointness of both the spatial part of wave operator and the
Hamiltonian operator of the one-dimensional Schr\"{o}dinger - like wave
equation. In our analysis, two different methods are used to test for
self-adjointness of the operators. The results of both methods are in
complete agreement.

\subsection{The Dirac fields}

The Newman-Penrose formalism will be used to find the Dirac fields
propagating in the background geometry of the naked singular KS spacetime.
We follow the formalism of Chandrasekhar \cite{28} and, hence, we shift the
signature of the metric (5) to $-2.$ The Chandrasekhar-Dirac (CD) equations
in Newman-Penrose formalism are given by

\begin{eqnarray}
\left( D+\epsilon -\rho \right) F_{1}+\left( \bar{\delta}+\pi -\alpha
\right) F_{2} &=&0, \\
\left( \Delta +\mu -\gamma \right) F_{2}+\left( \delta +\beta -\tau \right)
F_{1} &=&0,  \notag \\
\left( D+\bar{\epsilon}-\bar{\rho}\right) G_{2}-\left( \delta +\bar{\pi}-%
\bar{\alpha}\right) G_{1} &=&0,  \notag \\
\left( \Delta +\bar{\mu}-\bar{\gamma}\right) G_{1}-\left( \bar{\delta}+\bar{%
\beta}-\bar{\tau}\right) G_{2} &=&0,  \notag
\end{eqnarray}%
where $F_{1},F_{2},G_{1}$ and $G_{2}$ are the components of the wave
function, $\epsilon ,\rho ,\pi ,\alpha ,\mu ,\gamma ,\beta $ and $\tau $ are
the spin coefficients. The non-zero spin coefficients are given in Eq.(18).
The solution procedure of the set of CD equations (61) is exactly the same
as in the references \cite{29,30}. Thus, applying the same procedures, we
end up with a resulting one-dimensional Schr\"{o}dinger-like wave equation
with effective potential that governs the Dirac field,

\begin{gather}
\left( \frac{d^{2}}{dr_{\ast }^{2}}+k^{2}\right) Z_{\pm }=V_{\pm }Z_{\pm },
\\
V_{\pm }=\left[ \frac{f\lambda ^{2}}{r^{2}}\pm \lambda \frac{d}{dr_{\ast }}%
\left( \frac{\sqrt{f}}{r}\right) \right] .
\end{gather}%
In these equations, $Z_{\pm }=R_{1}\pm R_{2}$, represents the combination of
the two solutions of the CD equations and $\lambda $ denotes the
separability constant.

\subsubsection{Method 1: The von Neumann Criterion of Deficiency Indices}

In analogy with equation (22), the radial operator $A$ for the Dirac
equations can be written as,

\begin{equation*}
A=-\frac{d^{2}}{dr_{\ast }^{2}}+V_{\pm }.
\end{equation*}%
If we write the above operator in terms of the usual coordinates $r,$ by
using $\frac{d}{dr_{\ast }}=f\frac{d}{dr}$, we have

\begin{equation}
A=-\frac{d^{2}}{dr^{2}}-\frac{f^{^{\prime }}}{f}\frac{d}{dr}+\frac{1}{f^{2}}%
\left[ \frac{f\lambda ^{2}}{r^{2}}\pm \lambda f\frac{d}{dr}\left( \frac{%
\sqrt{f}}{r}\right) \right] .
\end{equation}

Our aim now is to investigate whether this radial part of the Dirac operator
is essentially self-adjoint or not. We do this by considering Eq.(27) and
counting the number of solutions that do not belong to Hilbert space. Thus,
Eq.(27) becomes

\begin{equation}
\left( \frac{d^{2}}{dr^{2}}+\frac{f^{^{\prime }}}{f}\frac{d}{dr}-\frac{1}{%
f^{2}}\left[ \frac{f\lambda ^{2}}{r^{2}}\pm \lambda f\frac{d}{dr}\left( 
\frac{\sqrt{f}}{r}\right) \right] \mp i\right) \psi (r)=0.
\end{equation}%
The solutions of (65) should be tested for square integrability over all
space $L^{2}\left( 0,\infty \right) .$ To do this, the behavior of (65),
near $r\rightarrow 0$ and $r\rightarrow \infty $ will be considered
separately in the following subsections.

\paragraph{The case of $r\rightarrow 0:$}

Note that when $r\rightarrow 0,$ the metric function transforms to (30) and
using (30) in equation (65) yields,%
\begin{equation}
\psi ^{\prime \prime }+\frac{1}{2r}\psi ^{\prime }+\frac{\sigma }{r^{3/2}}%
\psi =0
\end{equation}%
where $\sigma =\frac{\lambda \left( \lambda \pm 3/2\right) }{\sqrt{2}},$
whose solution is%
\begin{equation}
\psi (r)=C_{5}r^{1/4}J_{1}\left( ar^{1/4}\right) +C_{6}r^{1/4}N_{1}\left(
ar^{1/4}\right) .
\end{equation}%
in which $C_{5}$, $C_{6}$\ are integration constants and $a=4\sqrt{\sigma }.$
The square integrability is checked by using the definition of norm given in
Eq.(29), in the limiting case of the metric (5) when $r\rightarrow 0$. The
result of our analysis is that the solution when $r\rightarrow 0$ is square
integrable, because $\Vert \psi \Vert ^{2}<\infty ,$ indicating that the
solution (67) belongs to the Hilbert space.

\paragraph{The case of $r\rightarrow \infty $ :}

In the limiting case of $r\rightarrow \infty $, using the metric function
(35) in (65), gives%
\begin{equation}
\psi ^{\prime \prime }\pm i\psi =0,
\end{equation}%
and its solution is given by,

\begin{equation}
R(r)=C_{7}\sin \eta r+C_{8}\cos \eta r,
\end{equation}%
in which $\eta =\frac{1}{\sqrt{2}}\left( i\pm 1\right) ,$ and $C_{7},$ $%
C_{8} $ are the integration constants ( in general complex). The square
integrability is checked with the norm defined in Eq.(29) written for the
case $r\rightarrow \infty .$ The result is that the solution fails to
satisfy square integrability condition ($\Vert R\Vert ^{2}\rightarrow \infty 
$), and hence, does not belong to the Hilbert space.

In view of the analysis, there is one solution \ (near, $r\rightarrow \infty 
$ ) that does not belong to the Hilbert space in the entire space. As a
result, the spatial operator $A$, has a unique extension and it is said to
be essentially self- adjoint. And, the future time evolution of the Dirac
field can be predicted uniquely. Therefore, the naked singularity of the KS
spacetime remains quantum regular when probed with fermions ($spin-1/2$)
obeying the CD equations.

\subsubsection{Method 2: Weyl's limit circle - limit point criterion}

The CD equations have been written in one-dimensional Schr\"{o}dinger - like
wave equation and the effective potential is found to be as in equation
(63). Now, the behavior of the effective potential will be analyzed near $%
r\rightarrow \infty $ and $r\rightarrow 0,$ for self-adjointness of the
Hamiltonian operator $H=-\frac{d^{2}}{dr_{\ast }^{2}}+V_{\pm }.$

\paragraph{The case when $r\rightarrow \infty :$}

We have stated earlier that in the limit $r\rightarrow \infty ,$ the
tortoise coordinate approximates to $r_{\ast }\simeq r.$ The effective
potential (63) in the limit $r\rightarrow \infty $ can be written as,%
\begin{equation}
V_{\pm }\simeq \frac{\lambda ^{2}\mp \lambda }{r^{2}}.
\end{equation}%
This effective potential has been analyzed for both possible cases, namely, 
\begin{equation}
V_{+}=\frac{\lambda ^{2}-\lambda }{r^{2}}\text{ \ \ \ \ \ \ \ \ \ \ \ \ and
\ \ \ \ \ \ \ \ \ \ \ }V_{-}=\frac{\lambda ^{2}+\lambda }{r^{2}}.\text{\ \ }
\end{equation}%
If we define $M(r)=\frac{\lambda }{r^{2}},$ then $V_{+}\geq -\frac{\lambda }{%
r^{2}}.$ Thus, the requirement of item (ii) of Theorem 3, showed that, 
\begin{equation}
\int_{1}^{\infty }\left( M(r)\right) ^{-1/2}dr=\int_{1}^{\infty }\left( 
\frac{\lambda }{r^{2}}\right) ^{-1/2}dr=\infty
\end{equation}%
and hence, it is satisfied. Item (iii) of the Theorem 3, indicates that $%
\left( M(r)\right) ^{\prime }=-\frac{2\lambda }{r^{3}},$ and 
\begin{equation}
\frac{\left( M(r)\right) ^{\prime }}{\left( M(r)\right) ^{3/2}}=-\frac{%
2\lambda }{\lambda ^{3/2}}<\infty ,
\end{equation}%
which is bounded near infinity. Our analysis has indicated that for the
effective potential $V_{+},$ the Hamiltonian operator is in the limit point
case. However, the effective potential $V_{-}$ do not satisfy the
requirements of the Theorem 3 and hence, for this particular case the
Theorem 3 does not work. In order to clarify this case, we use the Corollary
defined in the Appendix.

When the first item of the Corollary is used as below, we have 
\begin{equation}
\int_{1}^{\infty }\frac{dr}{\sqrt{K-V_{\pm }(r)}}=\int_{1}^{\infty }\frac{dr%
}{\sqrt{K-\frac{\left( \lambda ^{2}\mp \lambda \right) }{r^{2}}}}=\frac{1}{K}%
\sqrt{Kr^{2}-\left( \lambda ^{2}\mp \lambda \right) }\left\vert _{1}^{\infty
}\right. =\infty ,
\end{equation}%
which satisfies the item $(i)$. \ The second item of the Corollary reveals
that%
\begin{equation}
\left( V_{\pm }(r)\right) ^{\prime }\left\vert V_{\pm }(r)\right\vert
^{-3/2}=-2\left( \lambda ^{2}\mp \lambda \right) \left\vert \lambda ^{2}\mp
\lambda \right\vert ^{-3/2}\text{ \ \ \ \ }
\end{equation}%
which is bounded near infinity. In view of this analysis, we can conclude
that the effective potential is in the limit point case.

\paragraph{The case when $r\rightarrow 0:$}

Near $r\rightarrow 0,$ the tortoise coordinate becomes $r_{\ast }\simeq 1+r,$
and the effective potential in terms of leading terms as $r\rightarrow 0,$
is given by%
\begin{equation}
V_{\pm }\simeq \frac{\lambda ^{2}\mp \lambda }{r^{2}}.
\end{equation}%
As a result, by Theorem 4, if $\left( \lambda ^{2}\mp \lambda \right) \geq
3/4,$ then the Hamiltonian operator is in the limit point case at zero and
therefore, it is essentially self-adjoint.

In view of this analysis, the Hamiltonian operator both at zero and infinity
is essentially self-adjoint. Hence, the KS naked singularity remains quantum
regular when probed with fermionic waves. Again, as in the case of bosonic
waves, the results of the two methods are in complete agreement.

\section{Conclusion and Discussion}

We have studied the KS naked singularity in Ho\v{r}ava's gravity in view of
quantum mechanics. In our analysis, we have used the HM criterion that
incorporates with the essential self-adjointness of the spatial part of the
wave operator $A$ in the natural Hilbert space of quantum mechanics. This
space is a linear function space with square-integrable functions $%
L^{2}\left( 0,\infty \right) .$

In our analysis, the KS naked singularity is probed with two different types
of quantum fields. Bosonic waves (scalar wave, with $spin-0$) and fermionic
waves (Dirac field, $spin-1/2$) governed by the Klein-Gordon and CD
equations, respectively, are used. The calculations have revealed that when
the singularity is probed with bosonic and fermionic waves, the spatial part
of the wave operator $A$ and the Hamiltonian operator of the one-dimensional
Schr\"{o}dinger - like wave equation on the KS naked singular spacetime is
essentially self-adjoint.

The essential self-adjointness in both probe implies that if quantum field
dynamics, in other words, waves are considered in place of classical
particles dynamics, i.e. geodesics, the KS naked singularity is
"smoothed-out". As a result, the classically KS naked singular spacetimes
becomes quantum mechanically regular.

The notable outcome of this study is that the quantum nature of the KS naked
singularity has a distinctive character when compared with its analog models
in classical general relativity. As it was demonstrated in \cite{22,23} the
naked singularities in negative mass Schwarzschild $(m<0),$ and the extremal
Reissner-Nordstr\"{o}m $(\left\vert e\right\vert >m)$ spacetimes were
quantum mechanically singular.

In a parallel study, the quantum nature of a quantum cosmological model
within the context of HL gravity is considered in \cite{31}. It was shown
that the quantum Friedmann-Lema\^{\i}tre-Robertson-Walker universe filled
with radiation in the context of HL gravity is quantum mechanically
nonsingular. The result in cosmological models and our findings for the KS
naked singular spacetimes show that it is possible to heal the apparent
singularities in HL gravity within the framework of quantum mechanics.

\section{Appendix}

The two widely used theorems for determining the essential self-adjointness
of the operator $A$ are briefly presented.

\subsection{The von Neumann Criterion of Deficiency Indices}

This method for determining the number of self-adjoint extensions of the
operator $A$ was discovered by Weyl \cite{26}, and generalized by von
Neumann \cite{27}. The deficiency subspaces $N_{\pm }$ are defined by

\begin{align}
N_{+}& =\{\psi \in D(A^{\ast }),\text{ \ \ \ \ \ \ }A^{\ast }\psi =Z_{+}\psi
,\text{ \ \ \ \ \ }ImZ_{+}>0\}\text{ \ \ \ \ \ with dimension }n_{+} 
\tag{A1} \\
N_{-}& =\{\psi \in D(A^{\ast }),\text{ \ \ \ \ \ \ }A^{\ast }\psi =Z_{-}\psi
,\text{ \ \ \ \ \ }ImZ_{-}<0\}\text{ \ \ \ \ \ with dimension }n_{-}  \notag
\end{align}%
The dimensions $\left( \text{ }n_{+},n_{-}\right) $ are the deficiency
indices of the operator $A$. The indices $n_{+}(n_{-})$ are completely
independent of the choice of $Z_{+}(Z_{-})$ depending only on whether or not 
$Z$ lies in the upper (lower) half complex plane. Generally one takes $%
Z_{+}=i\lambda $ and $Z_{-}=-i\lambda $ , where $\lambda $ is an arbitrary
positive constant necessary for dimensional reasons. The determination of
deficiency indices is then reduced to counting the number of solutions of $%
A^{\ast }\psi =Z\psi $ ; (for $\lambda =1$),

\begin{equation}
A^{\ast }\psi \pm i\psi =0.  \tag{A2}
\end{equation}%
that belong to the Hilbert space. 
\newtheorem{theorem1}{Theorem} 
\begin{theorem1}
For an operator A with deficiency indices $\left( \text{ }n_{+},n_{-}\right) 
$ there are three possibilities

$(i)$ If $n_{+}=n_{-}=0,$ then A is (essentially) self-adjoint (in fact, this
is a necessary and sufficient condition).

$(ii)$ If  $n_{+}=n_{-}=n\geq 1,$ then A has infinitely many self-adjoint
extensions, parametrized by a unitary $n\times n$ matrix.

$(iii)$ If $n_{+}\neq n_{-},$ then A has no self adjoint extension.
\end{theorem1}

In view of this theorem, if there are no square integrable solutions ( i.e. $%
n_{+}=n_{-}=0$ ) for all space $\left( 0,\infty \right) $, the operator $A$
possesses a unique self-adjoint extension and thus, it is essentially
self-adjoint.

\subsection{Weyl's limit circle - limit point criterion}

A theorem of Weyl \cite{26,35}, relates the essential self-adjointness of
the Hamiltonian operator to the behavior of the effective potential of the
one-dimensional Schr\"{o}dinger - like wave equation, which in turn
determines the behavior of the wave packet. This involves to determine
whether the effective potential is in the limit circle or limit point case.
The radial part of the wave equation can be written as a one-dimensional \
Schr\"{o}dinger - like equation $H\psi (x)=\lambda \psi (x)$ where the
Hamiltonian operator $H=-\frac{d^{2}}{dx^{2}}+V(x)$ and $\lambda $ is a
constant. Here, any singularity is assumed to be located at $x=0$. Reed and
Simon \cite{35}, states the following definition. 
\newtheorem*{definition}{Definition} 
\begin{definition}
The potential $V(x)$ is in the \textbf{limit circle case} at infinity
(respectively at zero) if for some, and therefore all, $\lambda ,$all
solutions of%
\begin{equation*}
-\frac{d^{2}\psi \left( x\right) }{dx^{2}}+V(x)\psi \left( x\right) =\lambda
\psi \left( x\right) 
\end{equation*}%
are square integrable at infinity (respectively at zero). If $V(x)$ is not
in the limit circle case at infinity (respectively at zero), it is said to
be in the \textbf{limit point case}.
\end{definition}
This definition clearly states that, whether the potential $V(x)$ is in the
limit circle case or in the limit point case indicates if the solutions to
the one-dimensional Schr\"{o}dinger - like wave equation are unique. There
are two linearly independent solutions at infinity ( respectively at zero )
for the Schr\"{o}dinger - like wave equation for a given $\lambda .$If $V(x)$
is in the limit circle case at infinity ( respectively at zero ), both
solutions are square integrable at infinity ( respectively at zero ), and
also, all linear combinations are square integrable as well. But if there is
one solution that fails to be square integrable then $V(x)$ is in the limit
point case. This is the main idea of testing for quantum singularities;
there is no singularity in quantum mechanical point of view if the solution
is unique, as it is in the limit point case \cite{32}. The following
theorems from \cite{35}, give us a criterion to decide whether the
Hamiltonian operator $H=-\frac{d^{2}}{dx^{2}}+V(x)$ is essentially self
adjoint (i.e. unique self adjoint extension) or not. 
\newtheorem{theorem2}[theorem1]{Theorem} 
\begin{theorem2}
\textbf{Weyl's limit point-limit circle criterion} (Theorem X.7 of Ref. \cite%
{35}). Let $V(x)$ be a continuous real-valued function on $\left( 0,\infty
\right) $. Then $H=-\frac{d^{2}}{dx^{2}}+V(x)$ is essentially self-adjoint
on $C_{0}^{\infty }\left( 0,\infty \right) $ if and only if $V(x)$ is in the
limit point case at both zero and infinity.
\end{theorem2}

At infinity $(x\rightarrow \infty ),$ the limit circle- limit point behavior
can be established with the help of the following theorem and its subsequent
corollary. 
\newtheorem{theorem3}[theorem1]{Theorem} 
\begin{theorem3}
(Theorem X.8 of Ref. \cite{35}). Let $V(x)$ be a continuous real-valued
function on $\left( 0,\infty \right) $ and suppose that there exists a
positive differentiable function $M(x)$ so that

$(i)$ $V(x)\geq -M(x)$

$(ii)$ $\int_{1}^{\infty }\left( M(x)\right) ^{-1/2}dx=\infty $

$(iii)$ $M^{\prime }(x)/\left( M(x)\right) ^{3/2}$ is bounded near $\infty .$

Then V(x) is in the limit point case (complete) at $\infty .$

\end{theorem3}
\newtheorem*{corollary}{Corollary} 
\begin{corollary}
(Corollary following Theorem X.8 of Ref. \cite{35}). Let $V(x)$ be differentiable on $\left( 0,\infty \right) $ and bounded above
by K on $\left[ 1,\infty \right) .$ Suppose that

$(i)$ $\int_{1}^{\infty }\frac{dx}{\sqrt{K-V(x)}}=\infty .$

$(ii)$ $\left( V(x)\right) ^{\prime }\left\vert V(x)\right\vert ^{-3/2}$ is
bounded near infinity. \ 

Then $V(x)$ is in the limit point case at $\infty .$

\end{corollary}
At zero $(x\rightarrow 0),$the limit circle- limit point behavior can be
established with the help of the following theorem. 
\newtheorem{theorem4}[theorem1]{Theorem} 
\begin{theorem4}
(Theorem X.10 of Ref. \cite{35}). Let $V(x)$ be a continuous and positive
near zero. If $V(x)\geq \frac{3}{4}x^{-2}$ near zero then $H=-\frac{d^{2}}{%
dx^{2}}+V(x)$ is in the limit point case at zero. If for some $\epsilon \geq
0,$ $V(x)\leq \left( \frac{3}{4}-\epsilon \right) x^{-2}$ near zero, then $%
H=-\frac{d^{2}}{dx^{2}}+V(x)$ is in the limit circle case.

\end{theorem4}

\textbf{Acknowledgment:}

We extend our most sincere gratitude to M. Halilsoy and S. H. Mazharimousavi
for their helpful comments.

\end{document}